\documentclass{blois}

\def\be{\begin{equation}}
\def\ee{\end{equation}}
\def\bea{\begin{eqnarray}}
\def\eea{\end{eqnarray}}

\newcommand{\x}{X(3872)}
\newcommand{\y}{Y(4260)}
\newcommand{\z}{Z_c(3900)}
\newcommand{\pp}{\pi^{+}\pi^{-}}

\newcommand{\EE}{e^{+}e^{-}}
\newcommand{\jpsi}{J/\psi}
\newcommand{\ppjpsi}{\pi^{+}\pi^{-}J/\psi}
\newcommand{\pphc}{\pi^{+}\pi^{-}h_c}

\begin{document}
\vspace*{4cm}
\title{$XYZ$ physics at BESIII experiment}

\author{ Zhiqing Liu }

\address{Johannes Gutenberg University of Mainz,\\
Johann-Joachim-Becher-Weg 45, D-55099 Mainz, Germany}

\maketitle\abstracts{
With the ability to run above 4~GeV, the BESIII experiment located in the Beijing Electron Positron Collider (BEPCII),
has becoming a pioneer in searching and studying charmoniumlike states ($XYZ$ particles). In 2013, BESIII Collaboration 
discovered a charged charmoniumlike state $\z$, which is confirmed immediately experimentally, and provides the best 
candidate for a four quark state by now. Continuous studies by BESIII Collaboration show new decay behavior of
$\z$, and there are possible partner particle $Z_c(4020)/Z_c(4025)$ existing. By scanning above 4~GeV, BESIII
also reveals the potential connection between $\y$ and $\x$ for the first time, which may help us
understand $XYZ$ particles in a new sight. 
}

\section{Introduction}

By decades, people know from the quark model~\cite{quark} that hadron matter existing in our universe
is composed of 3 quarks (baryon) or quark-anti-quark pairs (meson). However, QCD (the theory to describe strong 
force, which bind quarks together), allows new forms of matter in our universe, such as multi-quark states, 
hybrid states, glue balls and so on. Such kind of new hadrons are called exotic states.
On the other hand, we never saw an exotic hadron firmly in experiment. In the last ten years, 
many new particles in the charmonium (bound state of charm quark and anti-charm quark) mass region was observed. 
These new particles show different feature
with normal charmonium state, and might be good candidate for exotic states (they are called charmoniumlike states or
$XYZ$ particles). In 2003, the Belle experiment observed a new charmoniumlike state $\x$~\cite{x3872}, which is a
good candidate of four quark state. Later, the BaBar experiment observed another new charmoniumlike state $\y$~\cite{babary,belley},
which also might be exotic.

Although the process in charmoniumlike state is promising for evidence of exotic hadron, there is
still big difficulty in how to distinguish them from conventional charmonium states. However, this ambiguous situation changed
since 2013, when a new charged charmoniumlike state $\z$ discovered by the BESIII Collaboration~\cite{bes3-zc}, and immediately
confirmed by Belle Collaboration~\cite{belle-zc} and CLEO-c data~\cite{cleo-zc}. $\z$ carries electric charged, which
obvious can not be normal charmonium states. The minimal quark content of $\z$ should be four quark combination~\cite{view}.

The BESIII experiment is an $\EE$ machine running in the charmonium energy region. 
In this talk, we present the recent $XYZ$ physics results from the BESIII Collaboration, 
based on the high luminosity data sets collected above 4~GeV.

\section{Observation of $\z$ at BESIII}

The BESIII detector has collected 525~pb$^{-1}$ data at $\EE$ central-of-mass (CM) energy $(4.260\pm0.001)$~GeV. 
With this data sample, we analyze the $\EE\to\ppjpsi$ process~\cite{bes3-zc}. The Drift Chamber is used to catch 4 charged 
tracks ($\pp\ell^+\ell^-$), and the calorimeter is used to separate electrons and muons. 
We use the published Belle~\cite{belle-zc} and {\it BABAR}~\cite{babarnew} $\EE\to\ppjpsi$ cross section line shapes 
to do radiative correction. The Born order cross section at $\sqrt{s}=4.260$~GeV is measured to be  
$\sigma^{B}(\EE\to\ppjpsi)=(62.9\pm1.9\pm3.7)$~pb. The good agreement between BESIII, Belle~\cite{belle-zc} 
and {\it BABAR}~\cite{babarnew} for $\ppjpsi$ cross section measurement confirms the BESIII analysis is valid and unbiased.

\begin{figure}[htb]
\centering
\includegraphics[height=1.5in]{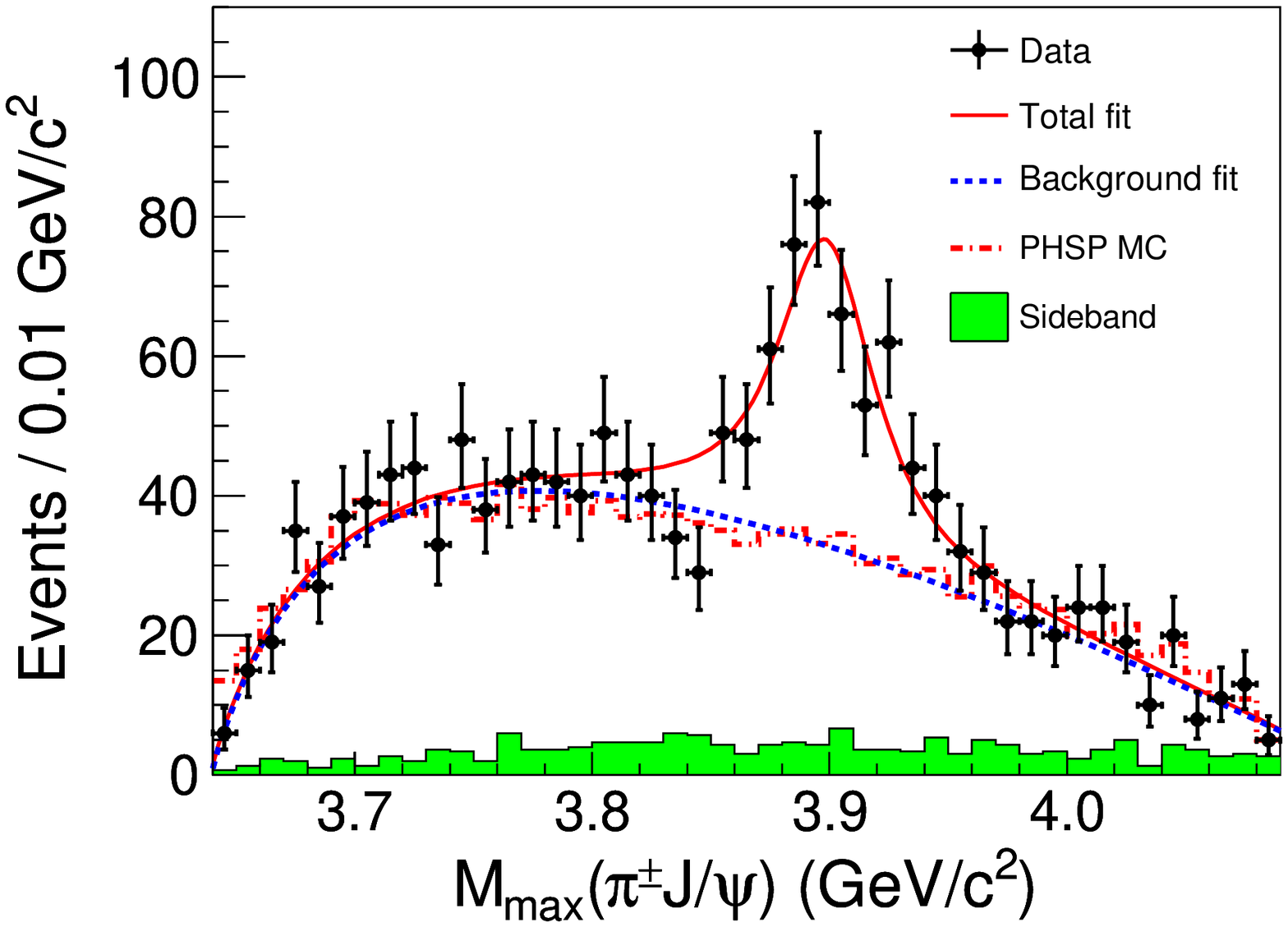}
\includegraphics[height=1.5in]{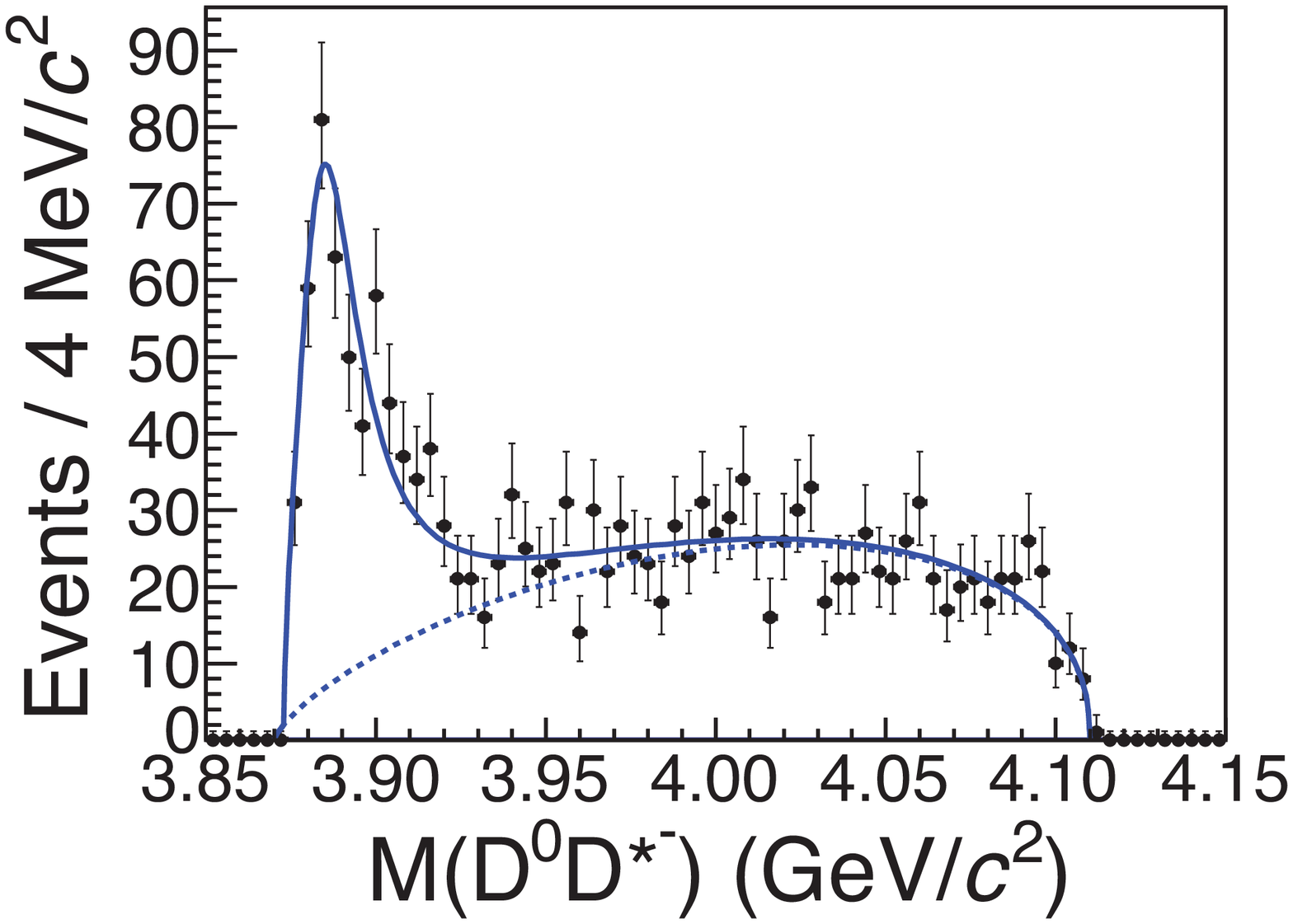}
\includegraphics[height=1.5in]{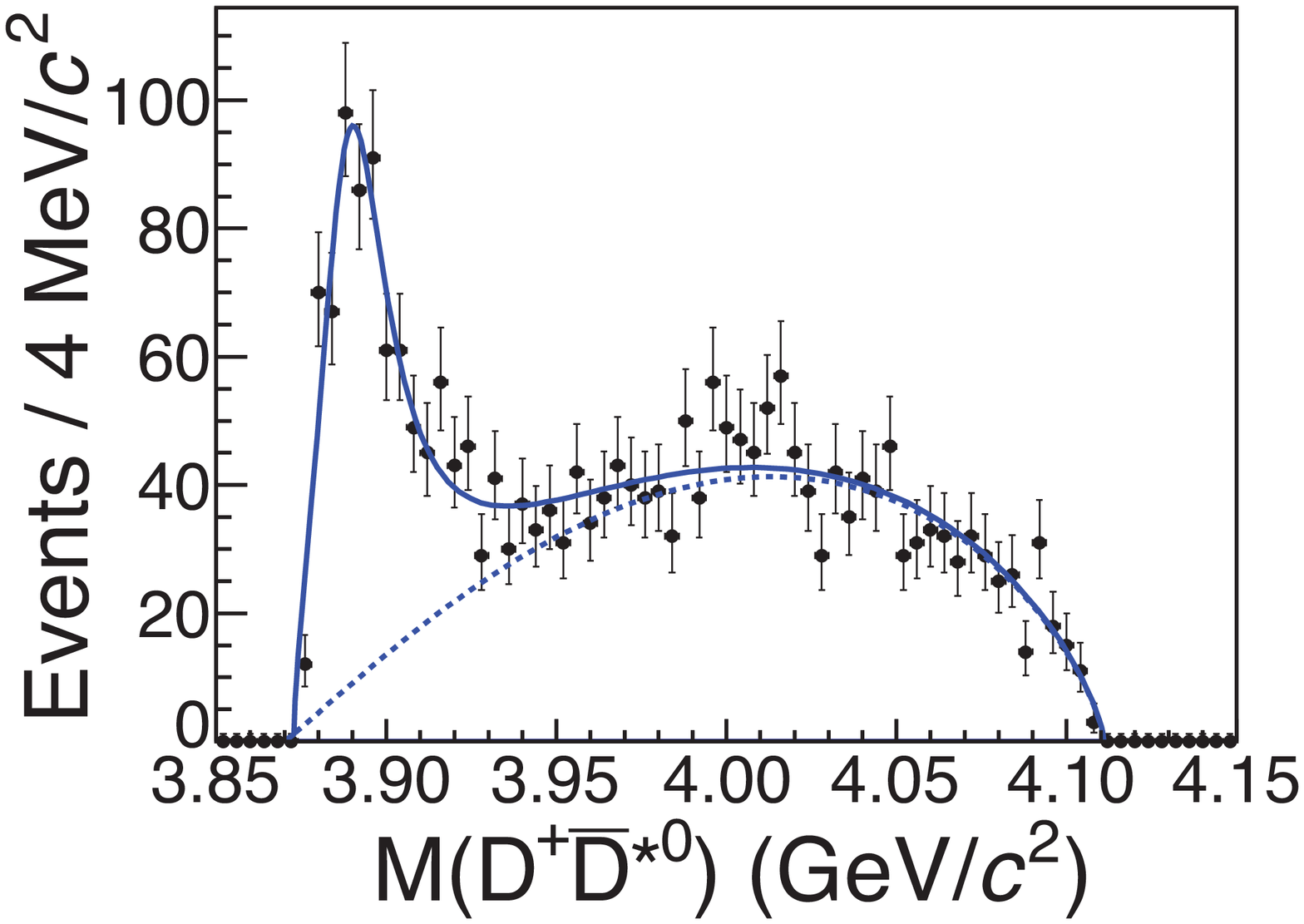}
\caption{Fit to the $M_{max}(\z\to\pi^\pm\jpsi)$ (left) and $M(\z\to D^0D^{*-})$ (middle) and $M(\z\to D^+\bar{D}^{*0})$ (right)  
invariant mass distribution as described in the text. Dots with error bars are data, the solid curves show the total fit and
the dashed curves are backgrounds contribution.}
\label{bes-fit}
\end{figure}

After obtained the cross section, we turn to investigate the intermediate state in $\y\to\ppjpsi$ decays. 
We got 1595 $\ppjpsi$ signal events with a purity of $\sim$90\%. 
The Dalizt plot of $\y\to\ppjpsi$ signal events shows interesting structures both in the $\pp$ system and 
$\pi^\pm\jpsi$ system. In the $\pi^\pm\jpsi$ mass distribution, a new resonance at around 
3.9~GeV/c$^2$ (called $\z$ hereafter) was observed. For the $\pp$ mass distribution, 
there are also interested structures, which can be modeled well by $0^{++}$ resonance $\sigma(500)$, $f_0(980)$ 
and non-resonant $S$-wave $\pp$ amplitude. The $D$-wave $\pp$ amplitude is found to be small in data 
and they also do not form peaks in the $M(\pi^\pm\jpsi)$ mass spectrum.
To extract the resonant parameters of $\z$, we use 1-dimensional unbinned maximum likelihood fit to the 
$M_{max}(\pi^\pm\jpsi)$ mass distribution (the larger one of $M(\pi^+\jpsi)$ and $M(\pi^-\jpsi)$ mass 
combination in each event), which is an effective way to avoid $\z^+$ and $\z^-$ components cross counting. 
Figure~\ref{bes-fit} (left) shows the fit results, with $M[\z]=(3899.0\pm3.6\pm4.9)$~MeV/c$^2$, and 
$\Gamma[\z]=(46\pm10\pm20)$~MeV. Here the first errors are statistical and the second systematic. 
The significance of $\z$ signal is estimated to be $>8\sigma$ in all kinds of systematic checks. 
%

\section{$\EE\to\pi^+(DD^*)^-$+c.c.}

The mass of $\z$ is a bit above $DD^*$ mass threshold, which motivates an assumption that $\z$ can coupling to $DD^*$. 
The BESIII Collaboration has performed the analysis of $\EE\to\pi^+(DD^*)^-$ (here charge
conjugation is always implied) with 525~pb$^{-1}$ data~\cite{zc3885}. The $(DD^*)^-$ system contains two combination: $D^0D^{*-}$ 
and $D^-D^{*0}$. In order to obtain more statistics, a good choice is to employed the partial reconstruction technique. 
The primary $\pi^+$ and $D$ meson are required to be detected, while the $D^*$ meson is missing.
The final 4-momentum of $DD^*$ system is obtained through $\EE$ initial momentum minus pion momentum,
which is due to strict momentum conservation. Figure~\ref{bes-fit} (middle, right) shows the obtained $DD^*$ 
invariant mass distributions. An obvious peak is observed near $DD^*$ mass threshold, which corresponds to a resonance. 
An unbinned maximum likelihood fit gives mass $M=3889.1\pm1.8$~MeV and width $\Gamma=28.1\pm4.1$~MeV 
($3891.8\pm1.8$~MeV and $27.8\pm3.9$~MeV) for the two data sets, respectively. The pole position of this peak is 
calculated to be $M_{\rm pole}=3883.9\pm1.5\pm4.2$~MeV and $\Gamma_{\rm pole}=24.8\pm3.3\pm11.0$~MeV,
where the first errors are statistical and the second systematic.

The mass and width of the peak observed in $DD^*$ final state agree with that of $\z$. Thus, they are quite probably the
same state. From the production cross section measurement, we also obtained 
$\frac{\Gamma[\z\to DD^*]}{\Gamma[\z\to\pi\jpsi]}=6.2\pm1.1_{\rm stat}\pm2.7_{\rm sys}$. That means $\z$ has a 
much stronger coupling to $DD^*$ than $\pi\jpsi$. Further study of production angle distribution shows the $DD^*$ peak
favor $J^P=1^+$ assignment.

\section{$Z_c(4020)$ and $Z_c(4025)$}

Using about 3.3~fb$^{-1}$ data, we also try to search charged charmoniumlike state in the 
$\EE\to\pphc$ process~\cite{zc4020}. The $h_c$ resonance is reconstructed through its radiative decay $h_c\to\gamma\eta_c$ 
(with $\sim50\%$ branching ratio), and $\eta_c$ resonance is reconstructed through 16 exclusive hadron decay channels
(with $\sim40\%$ branching ratios). After events selection, clear $\EE\to\pphc$ signal events are observed, and the $\EE$
CM energy dependent production cross section $\sigma^B(\EE\to\pphc)$ is measured, which is at the same order of that
$\ppjpsi$~\cite{bes3-zc,belle-zc,babarnew}.

\begin{figure}[htb]
\centering
\includegraphics[height=2.1in]{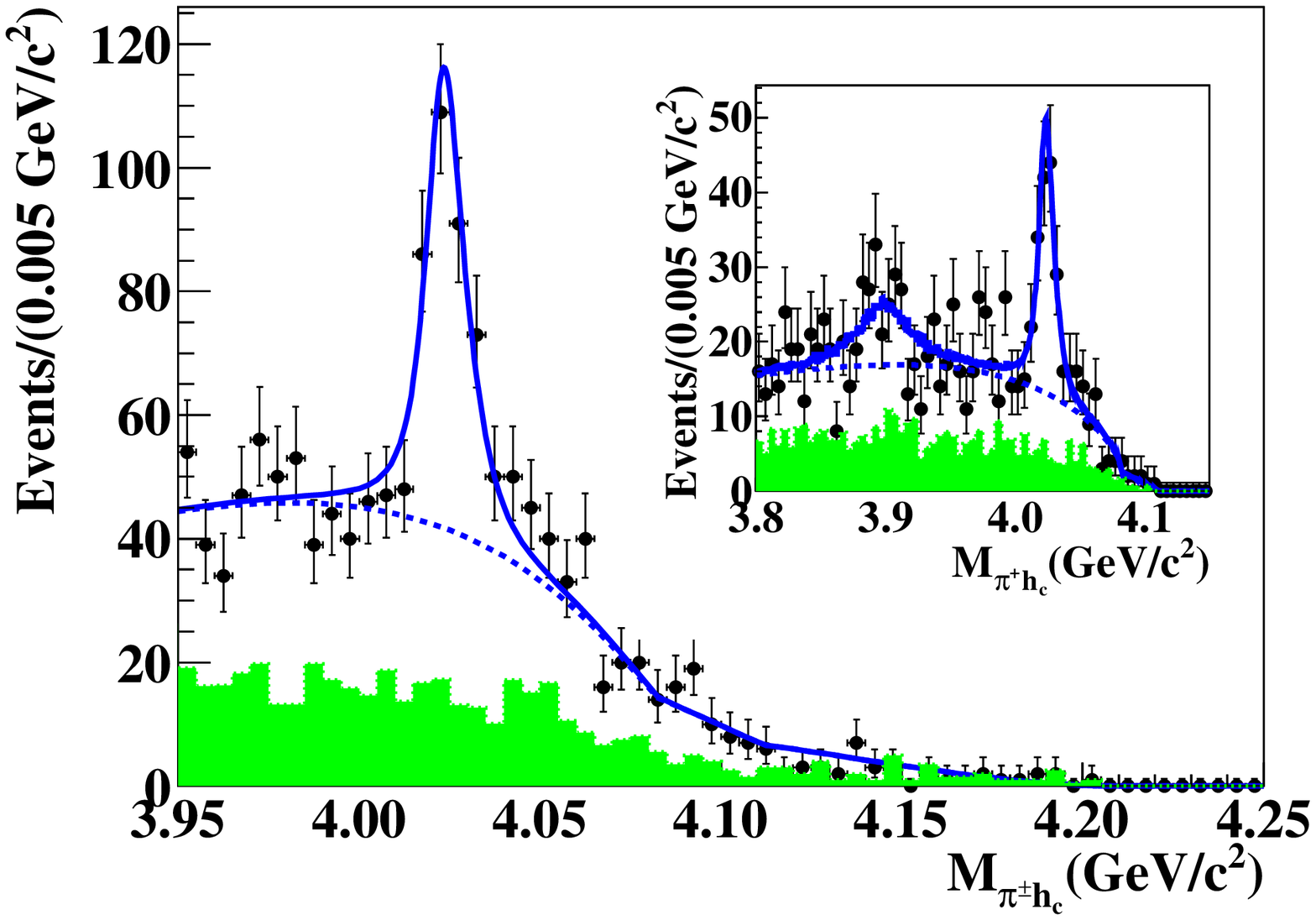}
\includegraphics[height=2.1in]{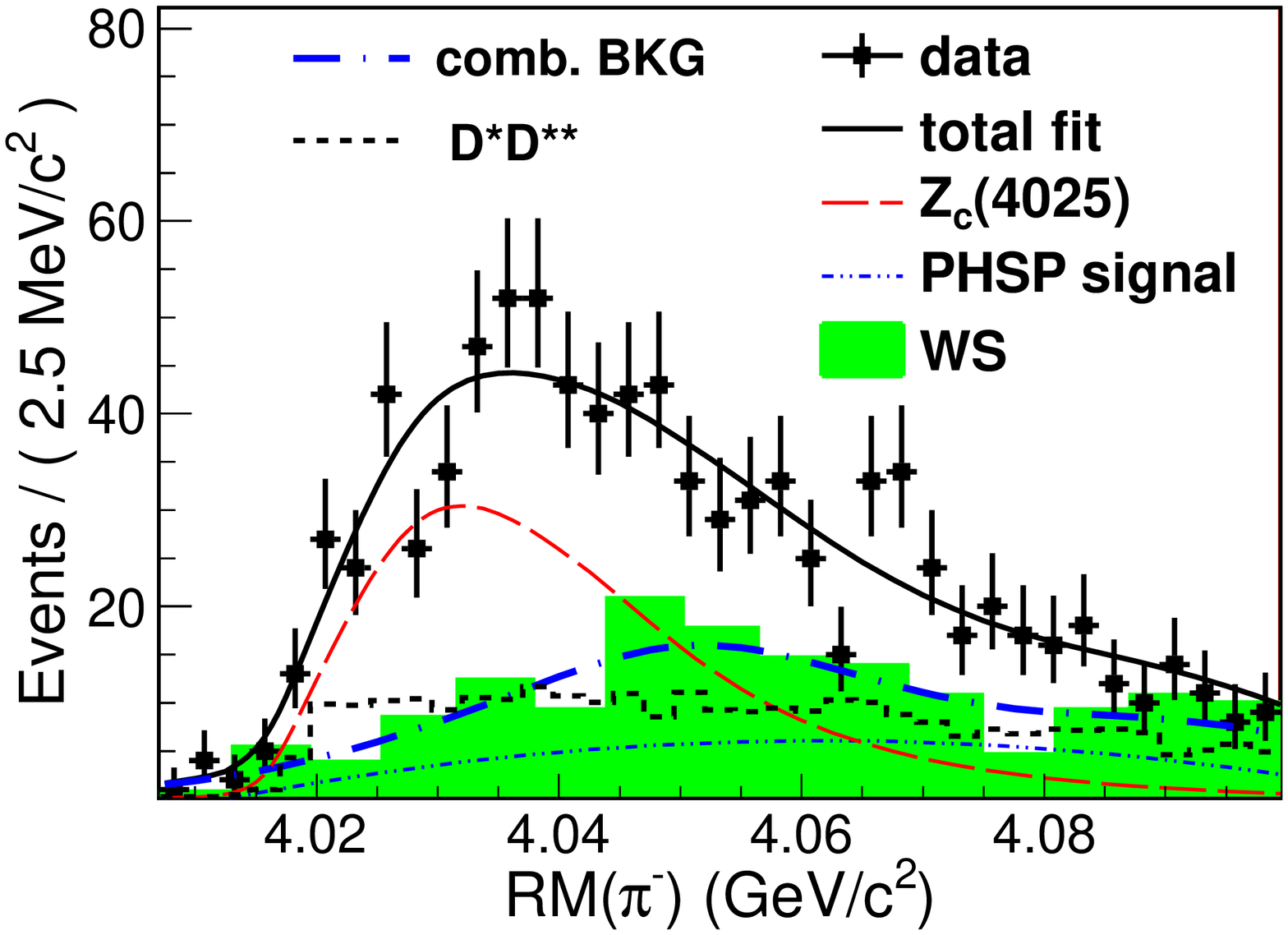}
\caption{$M(\pi^\pm h_c)$ (left) invariant mass distribution for $\EE\to\pphc$ data events 
and $M(D^*D^*)$ (right) invariant mass distribution for $\EE\to\pi^+(D^*D^*)^-$ data events.
Dots with error bars are data.
The solid curves in both panel are fit results, the dashed curve in left panel is background and in the
right panel is signal. }
\label{zcp}
\end{figure}

By further checking the $\pi^\pm h_c$ invariant mass distribution, a resonant structure was observed, as shown in Fig.~\ref{zcp} (left).
The measured mass is $M=4022.9\pm0.8\pm2.7$~MeV and width is $\Gamma=7.9\pm2.7\pm2.6$~MeV for the resonance (denoted
as $Z_c(4020)$), where the first errors are statistical and the second systematic. 
The significance of $Z_c(4020)$ is estimated to be $>8.9\sigma$. And the production cross section 
$\sigma^B(\EE\to\pi^+ Z_c(4020)^-\to\pphc)$ is measured to be $\sim10$~pb level at $\EE$ CM energy 4.23, 4.26, 4.36~GeV.
The $\z$ state is also searched, but find to be not significant in $\pphc$ process.

Being near $D^*D^*$ mass threshold, the $Z_c(4020)$ state is also quite possible to be coupling to $D^*D^*$. 
Using 827~pb$^{-1}$ data collected at $\sqrt{s}=4.26$~GeV, BESIII has studied the $\EE\to\pi^+(D^*D^*)^-$ (here
charge conjugation is always implied) process~\cite{zc4025}. In order to increase statistics, partial reconstruction technique is
also employed. The primary charged pion, one $D$ meson from charged $D^*$ decay, and at least one $\pi^0$
from $D^*$ decay are detected. The final 4-momentum of $(D^*D^*)$ system is determined from $\EE$ initial
momentum minus the primary pion momentum, due to strict momentum conservation.

Figure~\ref{zcp} (right) shows the $(D^*D^*)$ invariant mass distribution. There is obvious 
excess for data events distribution over background estimation. We assume this enhancement is due to a 
resonant structure, and labeled it as $Z_c(4025)$. 
The measured mass is $M=4026.3\pm2.6\pm3.7$~MeV, and width is
$\Gamma=24.8\pm5.6\pm7.7$~MeV, where the first errors are statistical and the second systematic.
The significance of $Z_c(4025)$ is estimated to be $13\sigma$.

Charged charmoniumlike states $Z_c(4020)$ and $Z_c(4025)$ show up with a similar mass (near $D^*D^*$ threshold). 
Thus, they might be the same resonance. If we assume so, we can measure the relative decay width of
$\frac{\Gamma[Z_c(4025)\to D^*D^*]}{\Gamma[Z_c(4020)\to\pi h_c]}\sim 9$. This behaves quite similar with
$\z$, and hints $Z_c(4020)/Z_c(4025)$ is a partner particle of $\z$.

\section{$\EE\to\gamma\x$}

The $\x$ was firstly observed by Belle Collaboration in $B\to K\ppjpsi$~\cite{x3872}. After ten years of its discovery,
its nature still keep mysterious. Recently, the LHCb Collaboration determined its quantum number to be $J^{PC}=1^{++}$~\cite{LHCb}.
Since BESIII can produce lots of vector particles $\psi/Y$s, thus it's natural to search for $\x$ in the radiative decay of vector particles.

\begin{figure}[htb]
\centering
\includegraphics[height=2.in]{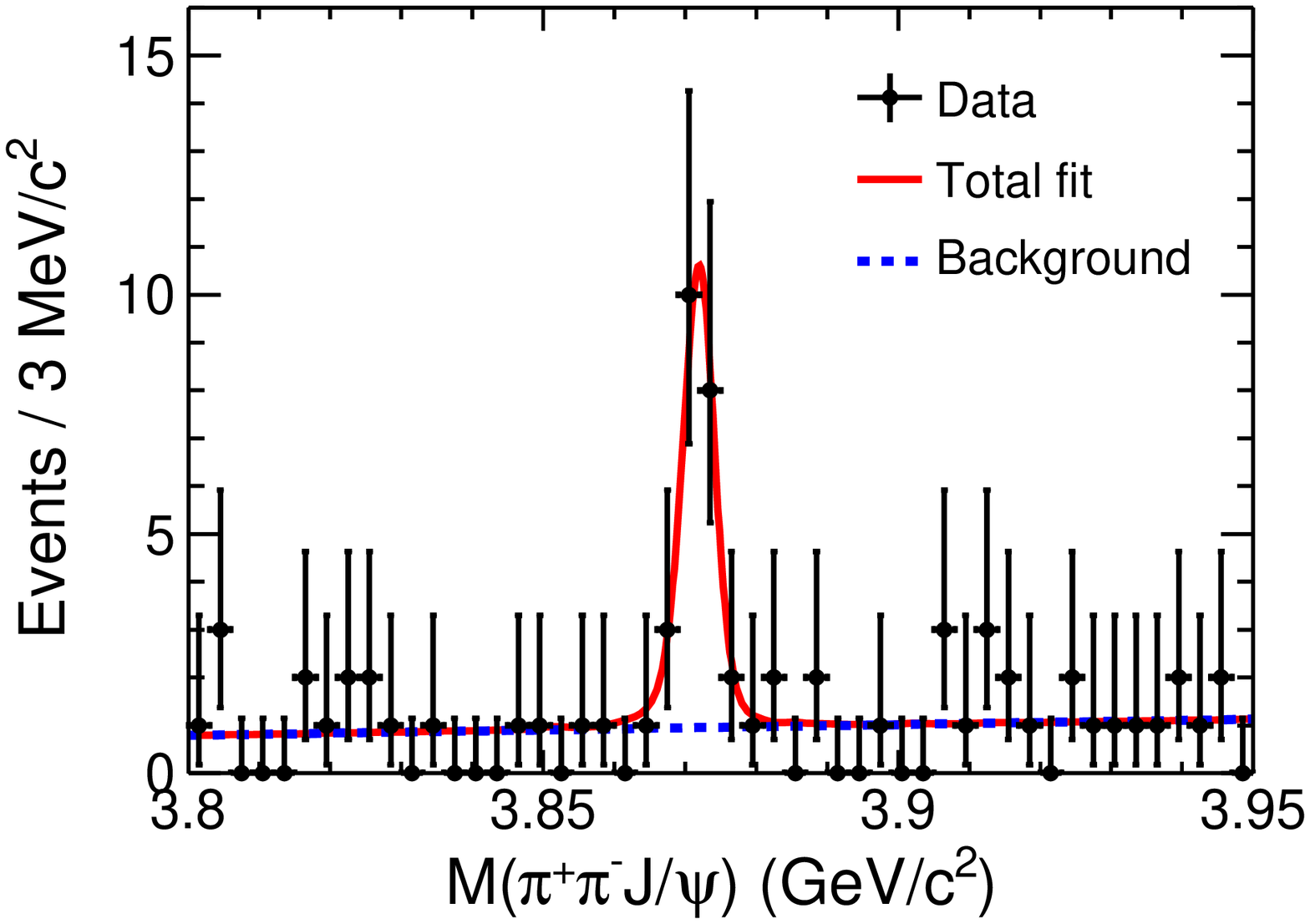}
\includegraphics[height=2.in]{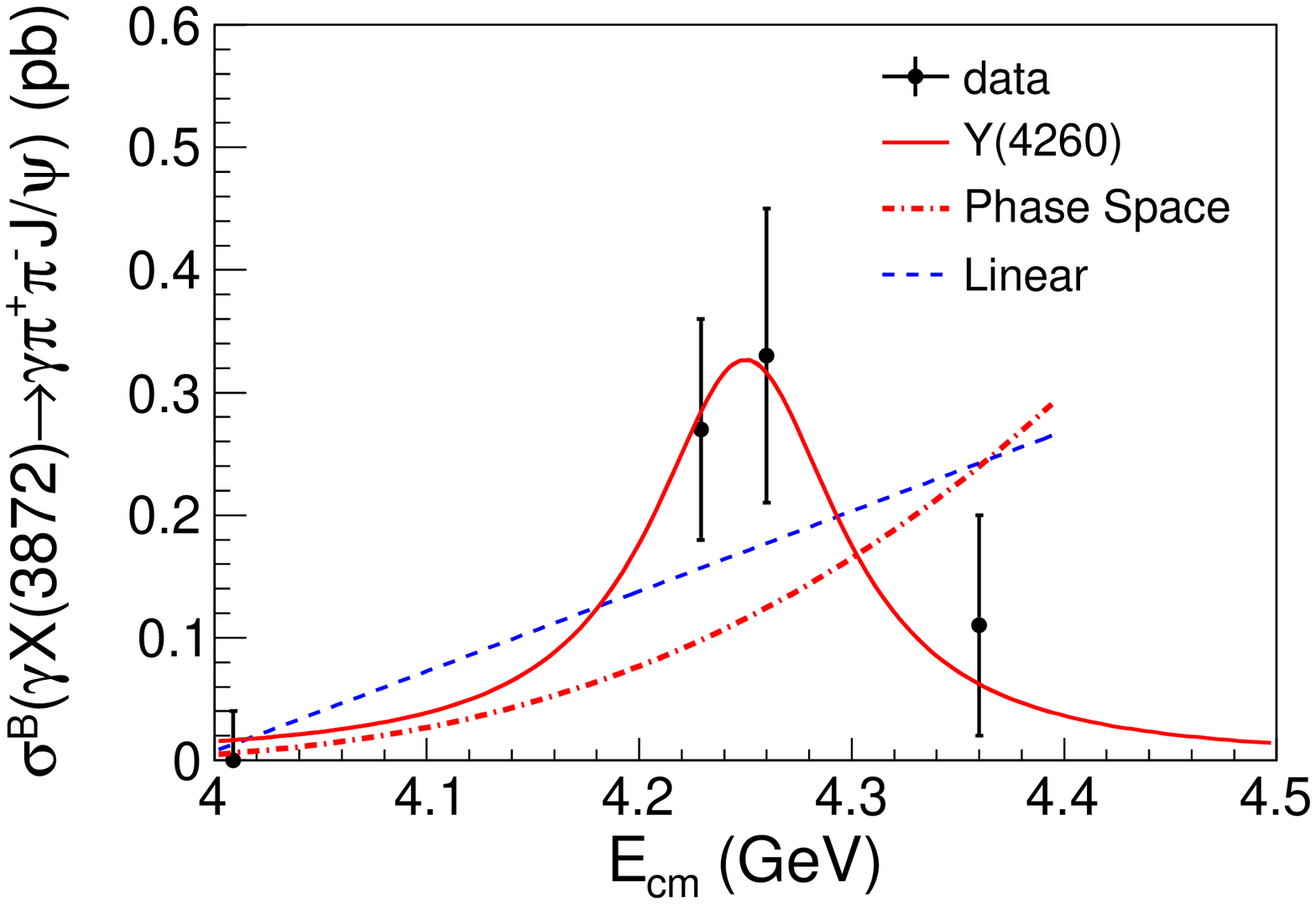}
\caption{$M(\ppjpsi)$ (left) invariant mass distribution for $\EE\to\gamma\ppjpsi$ events and
energy dependent cross section of $\EE\to\gamma\x$ (right). Dots with error bars are data.
In the left panel, the solid curve is fit, and the dashed one is background. In the right panel,
the curves show different fit results.}
\label{x-y}
\end{figure}

Using $\sim3.3$~fb$^{-1}$ data collected by BESIII, we have studied the $\EE\to\psi/Y\to\gamma\ppjpsi$ process~\cite{bes3-x3872}.
Figure~\ref{x-y} (left) shows the obtained $\ppjpsi$ invariant mass distribution from the whole data sets. $\x$
signal could be seen clearly. A fit to data events gives $M[\x]=3871.9\pm0.7_{\rm stat}\pm0.2_{\rm sys}$~MeV,
which agrees with other measurements very well~\cite{pdg}. The significant of $\x$ signal is estimated to be
$6.3\sigma$. It's worth to mention our measurement at BESIII provides another independent confirmation of
the $\x$ particle.

We also measured the $\EE$ CM energy dependent production cross section of $\gamma\x$. Figure~\ref{x-y} (right)
shows the cross section line shape, which peaks near 4.26~GeV. We find pure phase space and linear shape
describe the cross sections rather bad (with $\chi^2/ndf$=8.7/3 and 5.5/2, respectively), while $\y$ line shape can
describe the cross section line shape quite well (with $\chi^2/ndf=0.49/3$). It strongly suggested the decay
$\y\to\gamma\x$.


\section{Summary}

With the large data sets taken above 4~GeV, the BESIII experiment could study
$XYZ$ particles in a unique way. 
The charged charmoniumlike state $\z$ discovered recently by BESIII experiment 
give us solid evidence for an exotic hadron, probably a four quark state. 
Further study also shows $\z$ can couple to $DD^*$ final state strongly.
BESIII also observed a new charged charmoniumlike state $Z_c(4020)$, 
a ``partner" particle of $\z$. And a similar structure $Z_c(4025)$ (possible the same state as $Z_c(4020)$) 
was also found to be strongly coupling to $D^*D^*$. 

In addition to charged states, BESIII also studied $\x$ and $\y$ particles.
We observe the first radiative decay of $\y\to\gamma\x$, which connected
the $X$ and $Y$ particles together. Considering the $\z$ was also observed at $\sqrt{s}=4.26$~GeV,
it hints us there may be common nature for these $XYZ$ particles, and suggest
us understand them in a unified way. 

\section*{Acknowledgments}

This work was supported by the Marie Curie
International Incoming Fellowship within Seventh Framework 
Programme of the European Union under Grant Agreement
No. 627240.

\end{document}